# Evolution of science I: Evolution of Mind[1]

## Mayank N Vahia

Tata Institute of Fundamental Research, Mumbai

vahia@tifr.res.in

Abstract

The central nervous system and particularly the brain was designed to control the life cycle of a living being. With increasing size and sophistication, in mammals, the brain became capable of exercising significant control over life. In Homo Sapiens the brain became significantly powerful and capable of comprehension beyond survival needs with visualisation, formal thought and long term memory. Here we trace the rise of the powers of the brains of the Homo Sapiens and its capability of three comprehending the three spatial dimensions as well as time. By tracing the evolution of technology over the last millennium and particularly the late arrival of astronomy to discuss the evolution of the formal thinking process in humans. In a follow up paper we will trace the extensive use of this new faculty by humans to comprehend the working of the universe.

Introduction

Life exists in an environment where its primary purpose is to eat, not be eaten and to reproduce. In this eternal race for survival, size gives an obvious advantage. Dinosaurs took this idea to the extreme, building increasingly larger bodies. Eventually they became so large that they were incapable of adaptation and perished in a natural catastrophe.

The next best strategy was to make medium sized life forms that were more capable of adapting to different environments thus improving their chances of survival (see e.g. Bakker, 1971 regarding evolution of mammals and dinosaurs). This kind of formal adaptability was best served by a nervous system that was capable of understanding environment and responding to it in a complex manner to improve the survival chances of the life form. Such a nervous system required a flexible brain that needed to operate in a stable environment within the body of the life form. Due to the nature of biochemistry such a brain needed to work in a constant temperature environment. The optimum temperature for the working of the brain chemistry was slightly higher than the ambient temperature of the Earth in most environments. A series of adaptations eventually created warm blooded animals that could maintain a warm body and brain. These animals are classified as mammals and, in most cases, incubate their young ones inside their body to ensure that the brains of the young one grow well under the mother's warmth. They also look after the young during their early period after birth since these complex brains require post birth training to efficiently exploit their environment. While the earliest mammals existed during the time of dinosaurs (Kielan-jaworows et al., 2004) their room for evolution was heavily constrained by the giants on the ground.

Mammals flourished and became the most aggressive and dominant species[2] on the earth only after the disappearance of dinosaurs. The competition for increased adaptability was met through increased intelligence and more streamlined bodies. It eventually reached a stage where the homo series of animals, with the most efficient combination of body weight, adaptable body and a well-structured brain appeared on earth. This particular group of mammals had the highest brain to body ratio and ended up with intelligence in excess of what was needed for controlling the body and managing the immediate environment. They also had a versatile body form that permitted them to not only adapt better but also undertake systematic manipulation of their environment.

The mammals with the highest brain to body ratio, the Homo Sapiens appeared on earth about two million years ago and then further transformed into Homo Sapien Sapiens about six hundred thousand years ago (Guy et al., 2005, Gibbons 2013). The current body form that we call Homo Sapien Sapiens, at their peak, had a brain about one thousand five hundred cubic centimetre in volume about twenty thousand years ago. It seems to have been trimmed by nature to a more optimum size of one thousand three hundred and fifty cubic centimetre (about 10% reduction) at present apparently to optimize its function by optimising its design (Balter, 2002).

## 1. Evolution of human understanding of nature

Human investigations of nature and its manipulation go back about two million years probably with sticks and stones to acquire food. However, even the early Homo sapiens – the archaic humans – already had enough surplus computing power in their brains to go beyond simple adaptation to aggressive exploration of nature. Several early discoveries and inventions were made by archaic humans. These include flaked tools and hunting technologies (McBearty, 2012) and evidence of group hunting using spears 1.3 million years ago in Africa (Balter, 2010) and three hundred thousand years ago in Europe (Balter, 2014). They also spread over several parts of the earth. About six hundred thousand years ago one such group produced a variant with the highest brain to body ratio, firmly biped and a fast paced brain. These are called the Homo Sapien Sapiens. However, getting a perfect tuning of brain and body took a long time and adjusting various body parameters, including, for example, evolving vocal chords that could produce a large variety of complex sounds, came only about fifty thousand years ago (Balter, 2002). These are generally called behaviourally modern humans. But even before reaching this stage, humans had already invented jewellery about one hundred and thirty five thousand years ago (d'Errico et al., 2009), had learnt to control fire by three hundred thousand years ago (Roebroeks and Villa, 2011) and probably had a formal language by a hundred thousand years ago (Fitch, 2010). The oldest art on red ochre can be dated to a hundred thousand years ago (Henshilwood, 2011) with marked terracotta pieces going back to about seventeen thousand years (Henshilwood et al., 2002). The oldest human bed made with twigs dates back to about seventy seven thousand years ago (Wadley et al., 2011). They had begun to spread all over the globe by moving out of Africa about a hundred thousand years (Armitage et al., , 2011). Here they would meet other archaic humans and were capable of mating with them. There seem to have been at least two points in Europe and East Asia where they seem to have interbred (Vernot and Akey, 2014). In figure 1 below we give a brief timeline of various discoveries and inventions by humans. A broad outline of these developments is given in figure 1.

---

[2] We ignore the insects here since they ensure their survival more by group activity and a different response to the challenge of survival.

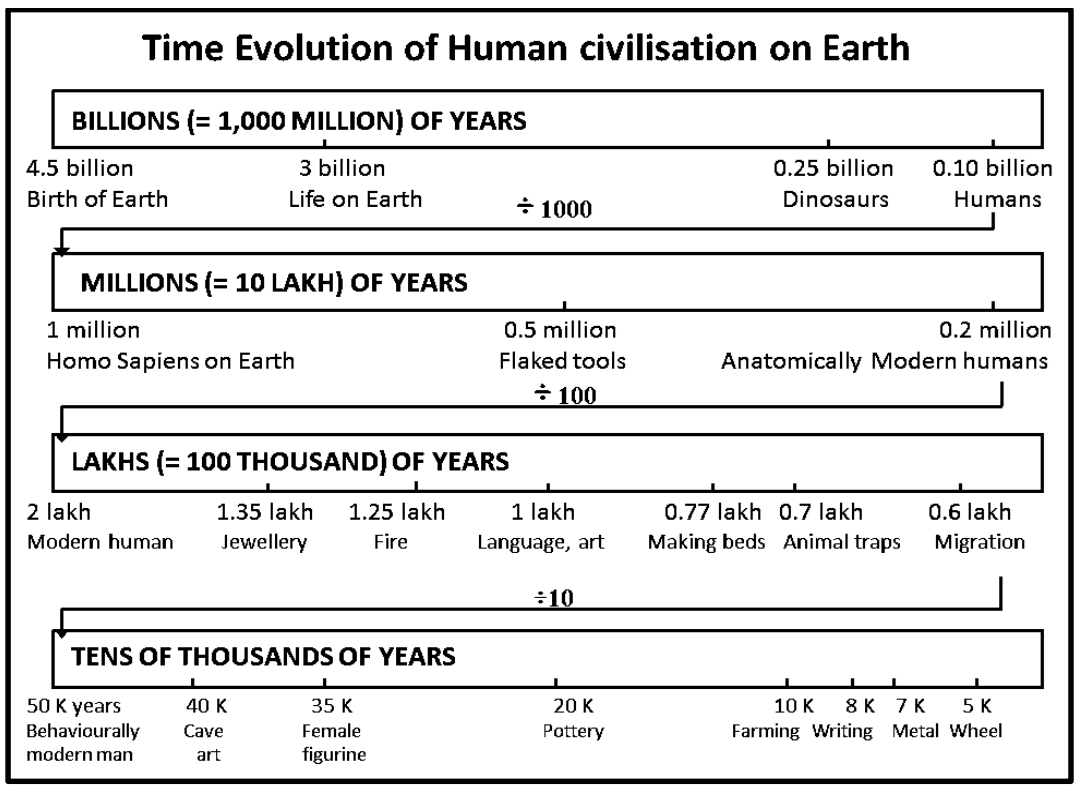

*Figure 1: Landmarks in development of technologies by humans*

## 2. Growth in human understanding of the environment

This progress was a result of significant developments in human comprehension of the universe. This is depicted in figure 2 below (see Vahia, 2014 for details). In the figure we quantify various aspects of human intelligences and environmental impact to study how the different faculties of human mind[3] evolved with time.

We begin with the minds of the great apes and study the evolution of the complexity in the perception with time (Figure 2). Several studies have shown the subtle but important differences between the minds of the great apes and humans, even as infants and children (see Tomasello, 2009 for a general summary). Similarly, studies such as Mithen (1996), Lewis-Williams (2002), Lewis-Williams and Pearce (2005) have considered the evolution of the Homo sapiens and early human minds from the perspective of growth and evolution of comprehension and early expressions of art. Based on these studies, it is clear that the world of the great apes is instinctive and they have only a weak concept of acquired knowledge or systematic collaboration of collation of knowledge over time. Here we extend these ideas to gauge specific aspects of evolution of the human brain. We divide the growth of human intellect from the period of great apes to humans in the following sequence.

---

[3] We use 'mind' to imply a mix of the hardware of the brain and the intellect that works with it.

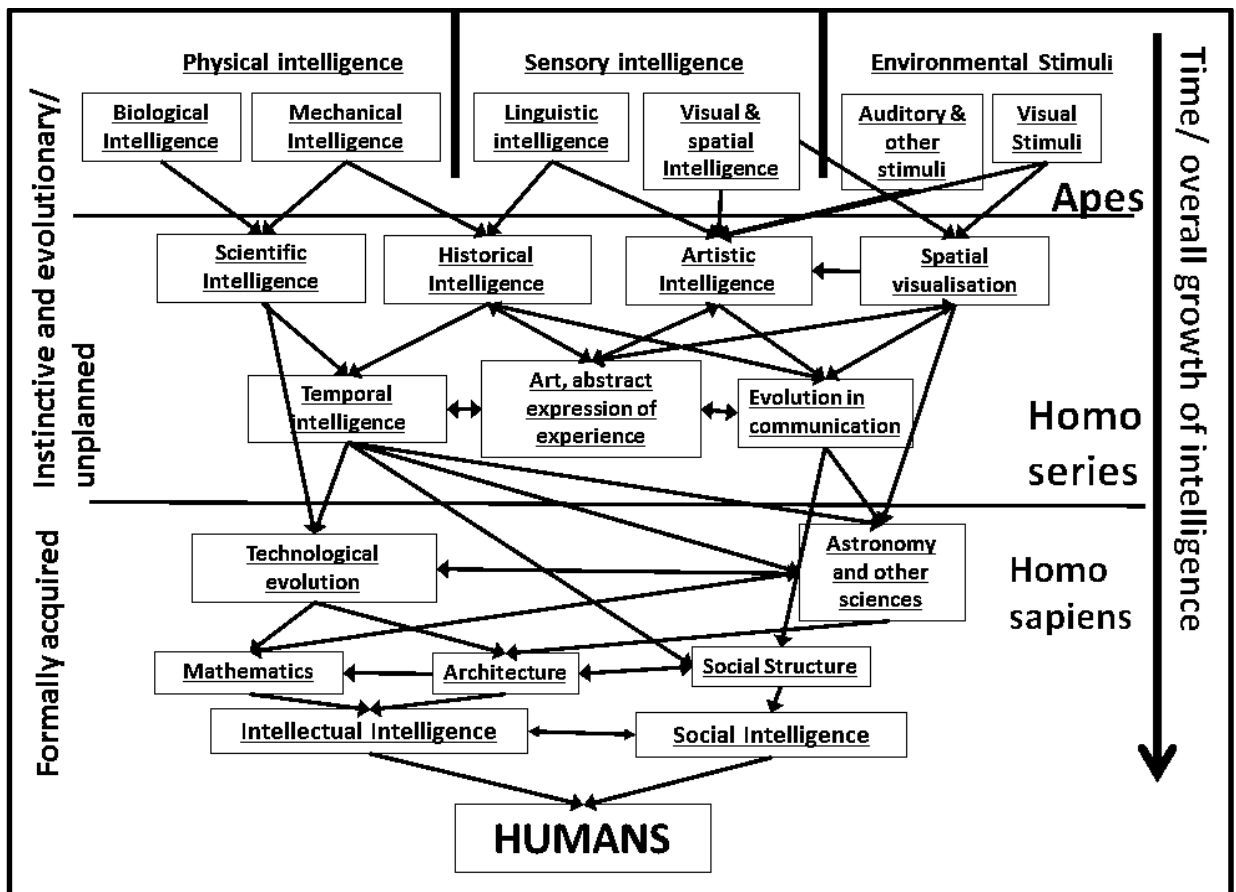

*Figure 2: Evolution of human intelligence with time*

In the natural environment, there are three forms of intelligence, namely physical intelligence, sensory intelligence and response to environmental stimuli. Physical intelligence in turn has two components, the biological and mechanical intelligence that is associated with the animate and inanimate world respectively. For this purpose the body has various sensory inputs that give the brain sensory intelligence consisting of visual and auditory intelligence etc. whereby it interprets the signals it receives. These arise from the stimuli it receives from the environment through its five senses. We distinguish between sensory intelligence and environmental stimuli in the sense that senses may exist and can work purely instinctively or may be interpreted to derive additional inferences. This requires an additional intelligence to interpret the signals conveyed by the senses received environmental stimuli.

With time, better organisation of the brain occurs, these acquire different features and the resultant intelligence can be divided into multiple components. It then derives a certain understanding about the physical world. In view of the nature of life, visual signals take up maximum brain power and have the highest priority followed by auditory signals. The sense of touch, taste and smell are primarily used for acquiring and processing the food. We classify it as biological Intelligence that deals with a variety of information. It becomes aware of body and response of the body to different stimuli, learns of manipulation of food for improving body response and use of plants for medicinal purpose, appreciates its body response and becomes sensitive to exhaustion, and illness. In humans and even in chimpanzees this results in becoming sensitive to the medical benefits of plants etc. (Huffman, 2003).

Mechanical intelligence arises from observing the properties of matter and manipulating them for tool making. The capacity of making controlled sounds and their recognition brings in linguistic intelligence where recognition of sound as an indication of occurrence of an unusual event including threats and treats is crucial. The auditory and olfactory senses sensitises the brain to inputs from sounds arising from animal movement and other natural phenomena, environmental stimuli as well as inter community and intra community signals. The inputs from touch and taste in turn allow sensitisation of the environment and provide a capability to understand immediate environment and its effects on the body. The visual and spatial intelligence provides a map of the environment and perception of motion also provides information on static environment, dynamic environment, signals from the environment in terms of sensitivity to the time of day, recognition of local landmarks. It can also invite hallucinatory stimuli with visualisation of abstract images. Lastly, when this information is processed by the brain in the sleep mode, it results in dreams.

We share most of these features with other primates and even lower animals. However, their complexity and sophistication vary. As the size of the brain and its internal organisation improves, the brain makes a transition from instinctive and evolutionary or unplanned working, to a more formal acquisition of knowledge. This acquisition of social identity, altruism and collaborative aspects seem to be pre-programmed in the human brain (Tomasello, 2009). Based on these stimuli, human brains with higher complexity and computing power will develop a whole series of complex intelligences. This will include an instinctive perception of the working of the environment that we call scientific intelligence which includes comprehension of the rules of nature, evolution concept of medical plants and repair procedure. The brain will also develop historical intelligence based on memory of past experiences. This will include comprehension from visual and other direct signals from predator/prey, recall of past experiences, recall of local geography and recall of common experience of the group through common experience.

Humans are also capable of several biological and mental gymnastics. These include use of opposable fingers to hold and manipulate objects, complex vocal chords to produce a large variety of sounds, a brain that can not only comprehend and store information but can also extrapolate to understand and anticipate actions of others based on minimal signals. Similarly, being biped gave humans a longer reach and upper limbs to dedicate to activities beyond locomotion. Together they give the humans a complex ability to be expressive beyond their survival needs. They can comprehend the working of nature and develop artistic intelligence to express themselves. This included comprehension of visual or auditory signals and reproduction complex of auditory signals and its subsequent analysis that allows humans to create a complex worldview that spanned long timescales. They can go beyond this through abstraction of visual signals and reproduce these visual signals through representative and performing arts. Under the correct conditions they can also hallucinate, go into trance and experience surrealistic experiences. By the same token they can also experience love and a sense of loss and grief at the loss of that love, be it for another human or an environment. Together this set of capabilities gives humans a strong sense of power and a desire as well as capability to manipulate their environment.

We have also paid a heavy evolutionary price for this. The physical development of a human young takes far longer than that of other animals. This has been attributed to the excessive metabolic demands of the brain which seems to suck away a large fraction of resource from human infants in the early years (Kuzawa et al., 2014). We also sustain a biped body that is clearly not fully efficient. The very fact that we need to change postures – sitting and sleeping horizontally – suggest that the muscle development and support system for the unstable equilibrium of biped style has not been fully balanced by our large flat feet and strong leg muscles which are reasonably adequate but are unable to sustain a standing posture over long periods of time. In other ways also the brain extracts a huge fraction of resources from the body (Vahia, 2014).

With increasing time evolution, two more features become apparent. The first, the complex temporal intelligence includes understanding of long term variation as well as time dependant changes, understanding tempo-spatial consequences of movement, a sense of events that occur over extended time scale and even notice their periodicity. With accumulation of historical information they even develop understanding of events over much larger time scales of human lifetime and beyond. Evolution in communication would lead to the development of language and recognition of emotions in animate sounds.

Being social animals, they also develop a complex work view which include a sense of evolution of time on different time scales to a visualisation of partially visible objects in static and dynamic environment through conscious and learned visualisation. With memory they can also identify periodic and a periodic event. It also forces them to create a series of ad hoc models to classify their information and convert it into a framework against which they keep track of the massive information and prevent overload of raw data. As intelligence evolves this becomes increasingly more sophisticated and complex. These frameworks can include both mythos and lagos the mythological and logical models to help them understand and predict the events in life. They can even create a mental image of the thought process of another being beyond immediate threat perception through what is called the theory of mind (Frith and Frith, 2005; Premack and Woodruff, 1978).

These features were probably common even before the modern humans, the Homo Sapiens, arrived on the earth. As a result of this complex evolution of human brain size and consequent improvement in computational capabilities, human mind would evolve even further in the form of technological evolution, namely, ability to manipulate environment and physical properties of material, creation of complex devices for efficient machines, manipulation of material property and modification of mechanical objects for specific purpose.

Such a learning would be backed up by a certain amount of social evolution and interactions including idea of self and perception of other individual's mind, development of collaboration and supportive altruism. They would also create myths and religion to try and give some form of explanation to the cause of the events around them. They would eventually create formal art and writing and recall from acquired knowledge, largely through language.

As a result of these developments, humans acquire a 4 dimensional world view – three of space and one of time. The realisation of the third dimension of space would introduce them to astronomy. This would open a completely new set of intellectual challenges from predicting seasons and calendar and a sense of vast sky that they would try and comprehend based on their world view (Vahia, Yadav and Menon, 2015).

The final evolutionary trait would recognise the two modern classification of human intelligence. This would involve what we call intellectual intelligence. This would include, intelligence of taste – nutrition relation, Intelligence to handle multi nodal inputs from different intelligences, specific search for information from a particular region. This in turn would provide them with an ability to mix comprehension and acquired experience an ability to control instinct, and involuntary action to perform unusual functions such as fainting as a defensive mechanism. Such a brain would be able to handle multi-nodal cross talk between senses and memory. It would also have a formal understanding of arithmetic and number system including formal capability of mathematical problem solving. The result would be an integrated understanding of physical sciences (physics, chemistry, biology etc. as well as mathematics). The brain would be in a position to comprehend of artificial signals, their visualisation and reality altering experiences such as those experienced from music and art. It would also be capable of creating of complex devices for efficient machines and manipulation of material property to create instruments such as animal traps and custom designed weapons and tools for individual needs and train younger generation to use this acquired knowledge that can significantly enhance human capabilities. The result would be complex interplay of developed technology that further increases understanding. The expanding capability would also be visible in abstraction and imaginative estimation of time and space evolved events and an informal recall of acquired knowledge largely through language and history with a sense of very long time scales beyond human existence.

Social intelligence would lead to development of features like humour with all its complexities (Hurley, et al., 2011). A sense of ethics, right and wrong and evaluation of consequences of preconceived action would result in evolution of social norms. Concepts of property and ownership would create its own sense of identity and community. In terms of common memory myths and religion would become formalised.

Expressions of thought would result in formal art and complex linguistic structures and grammar. Speculations on human relation to the environment would give the first philosophy and eventually writing would arrive as an aid to memory and then take its own separate path as a medium of self-expression. Ideas of history and philosophy would evolve. That cultural diversity and behavioural traits including choice of phonetics in a language (see below) is closely related the genetic polymorphism. This is well documented (Dediu and Ladd, 2007; Nettle, 2007).

The interrelations between them are highlighted in figure 2. Figure 2 has some interesting consequences. In particular, for the subjects such as astronomy to evolve, it is important that not only are the visual stimuli understood and analysed in terms of their complexity, but the fact that they vary with time also need to be understood before humans could appreciate astronomy. This can be seen from the plotting of points to indicate stars have been recorded in Lexus caves about 15000 years ago. In reality astronomical drawings are rare and astronomically aligned structures arise only around 3000 BC and earlier megalithic structures and residing caves give no indications that humans looked and appreciated the sky.

Another aspect of this growth is the fact that there is no real increase in the brain size of the Homo Sapiens for the last two hundred thousand years and probably a small reduction in size but an increase in the level of sophistication of the connectivity within the brain, and an indication of the increasing complexity of the brain's interconnectivity. The credit for this must go to the better organising capability that our leaner brain. These changes in the brain organisation and its plasticity  get transferred to later generation probably through genetic transfer indicating that the genes probably pre-decide many of the wirings in the brain at birth itself (see e.g. Dediu and Ladd, 2007 see also Nettle, 2007). Hence, while some of the jumps such as rise in the idea of altruism (Tomasello, 2009), can be attributed to the larger brains with increasingly complex connectivity, rise of perception in the Homo Sapiens and a natural adaptation must be attributed to acquired capabilities which we refer to as mind in further discussion. The fact that astronomy was not known to early Homo Sapiens with same brain size as ours indicates that they lacked the sense of movement of time and evolution of the 3 dimensional space the way we take as obvious. The fact that the even young children below 1 year seem to be capable of this comprehension of time and space movement suggests that there must have been significant evolution of the genetic markers related to brain that ensured that even with the same brain size, its connectivity (or the mind) continued to improve. This raises important questions about the feedback loop between social evolution and genetics.

Feigenson and Halberda (2008) have used 14 months' infants whose brains are not filled with meta-memorial strategies and shown that they can use their memories with complex subdivision of pattern to be memorised indicating that the fundamental memory computation is available before the brain adapts to meta-memorial strategies. While Feigenson and Halberda (2008) have used infant models, such studies have been used to understand time evolution of human brain capabilities. It is therefore well within the realm of possibility that after the Homo Sapiens acquired their large brain, improvement of interconnections started before language acquisition and was accelerated with the arrival of language.

Also, at least some of these developments seem to emerge at different locations nearly simultaneously, while others were clearly local in origin (Diamond, 2005). The general nature of this evolution can be seen from the fact that even though humans moved out from Africa well over 100,000 years ago, all major steps of growth of human cultures such as farming, writing, urbanisation etc. arrive in differ parts of the world around the same time even in diversely different environments and in a disconnected manner across several parts of the globe with similar compulsions. Ideas seem to arise in a natural sequence of smoothly increasing sophistication of the mind seems to suggest itself. At the same time, there are many significant differences which can be attributed to factors such as the lack of a particular technological advancement. Others can be attributed due to some ideas not arising in a particular cultural group. These differences then mark the differences in different cultures.

All these improvements in the brain then give rise to a realisation of the universe that is beyond senses, an independent entity and set of universal laws that are independent of the observer, the existence of a physical universe. The study of this universe is what we call science.

3. Rise in human comprehension and its impact on civilisation: A case study of astronomy

The rise in human intelligence is related to the complexity in a circular manner. The more complex the culture, the more sophistication it demanded from the brain, the more sophisticated the brain became, the more it learns to automate tasks, specialise itself and the more the society progressed. This progress of the society can be quantified in terms of various social parameters. This interplay has created truly complex organic cities where the interaction of individual intelligence, collective organisation, automation and environment have created a complex entity of its own right (Batty, 2013).

Astronomy generally begins with the belief in Mother Earth that needs to be seeded by Father Sky and the earliest image of a female with exaggerated femininity have been dated to thirty five thousand years (Conard, 2009). However, from this early approach to sky and the concept of Father in space, the idea of changing location of sunrise point over a year and constellations etc. is a large and complex development and is stretched over time. These include development of gnomons and sundials, megalithic horizon markers, constellation designs and associated developments of astronomical myths (Vahia, Yadav and Menon, 2014). Hence the recorded astronomical knowledge in each culture can be used to study its intellectual level. These growth in ideas can be quantified and studied as a function of other parameters of social sophistication. We use the period of stable settlement as a measure of the sophistication of their living conditions and note their astronomical knowledge that has been derived from archaeological and other studies to understand the relation between social complexity and study of nature.

We use the knowledge of astronomy as the indicator of their intellectual sophistication. In table 1 we list the period of settlement of various cultures and their level of their knowledge various aspect of astronomy. In figure 3 we plot a graph of the time since settlement of a culture against the level of complexity as derived from table 1.

| 20Community | Period since settlement (period since settlement when study is reported) | Level of astronomy | | | | | | | | | | Cumulative score | References |
|---|---|---|---|---|---|---|---|---|---|---|---|---|---|
| | | Name for Directions | Know constellations: <5 = 0.3; 5-10=0.6; >10=1 | Astronomy in daily life | Myths of seasons | Philosophical myths | Astrology | Eclipses | Planets | Lunar mansions | Observatories | | |
| Relative weight | | 1 | 10 | 5 | 8 | 10 | 10 | 10 | 10 | 10 | 10 | | |
| Gonds | 1000 | 1 | 0.6 | 0.5 | 1 | 1 | | | | | | 27.5 | Vahia and Halkare (2013) |
| Kolams | 500 | 1 | 0.6 | 0.5 | 1 | | 0.5 | 1 | | | | 32.5 | Vahia et al. (2014) |
| Banjaras | 100 | 1 | 0.3 | | 1 | | | | | | | 12 | Vahia et al. (2014) |
| Sumerians | 3000 | 1 | 1 | 0.5 | 1 | 1 | 1 | 1 | 1 | | 0.5 | 66.5 | |
| Greeks | 1000 | 1 | 0.3 | 0.5 | 1 | 1 | 1 | 1 | | | 0.5 | 49.5 | |
| Chinese (early period) | 3000 | 1 | 0.3 | 1 | 1 | 1 | 1 | 1 | 1 | 0.5 | 0.5 | 67 | |
| Egyptians | 1000 | 1 | 0.3 | 1 | 1 | 1 | 1 | | | | | 37 | |
| Greko Roman | 3000 | 1 | 1 | 1 | 1 | 1 | 1 | 1 | 1 | | 0.5 | 69 | Taken from Sumerians |
| Mayan | 4000 | 1 | 1 | 1 | 1 | 1 | 1 | 1 | 1 | 1 | 1 | 84 | |
| Rig Vedic | 1500 | 1 | 0.6 | 1 | 1 | 1 | 1 | 1 | | 1 | | 50 | Kuppanna Sastry (1984) |
| Late Vedic | 2000 | 1 | 1 | 1 | | 1 | | 1 | 1 | 1 | | 56 | Iyengar (2013) |
| India 500 AD | 3500 | 1 | 1 | 1 | 1 | | 1 | 1 | 1 | 1 | 0.5 | 79 | Balachandra Rao (2008) |
| Early UK | 1000 | 1 | 0.6 | 1 | 1 | | 1 | 1 | | 1 | 1 | 50 | |

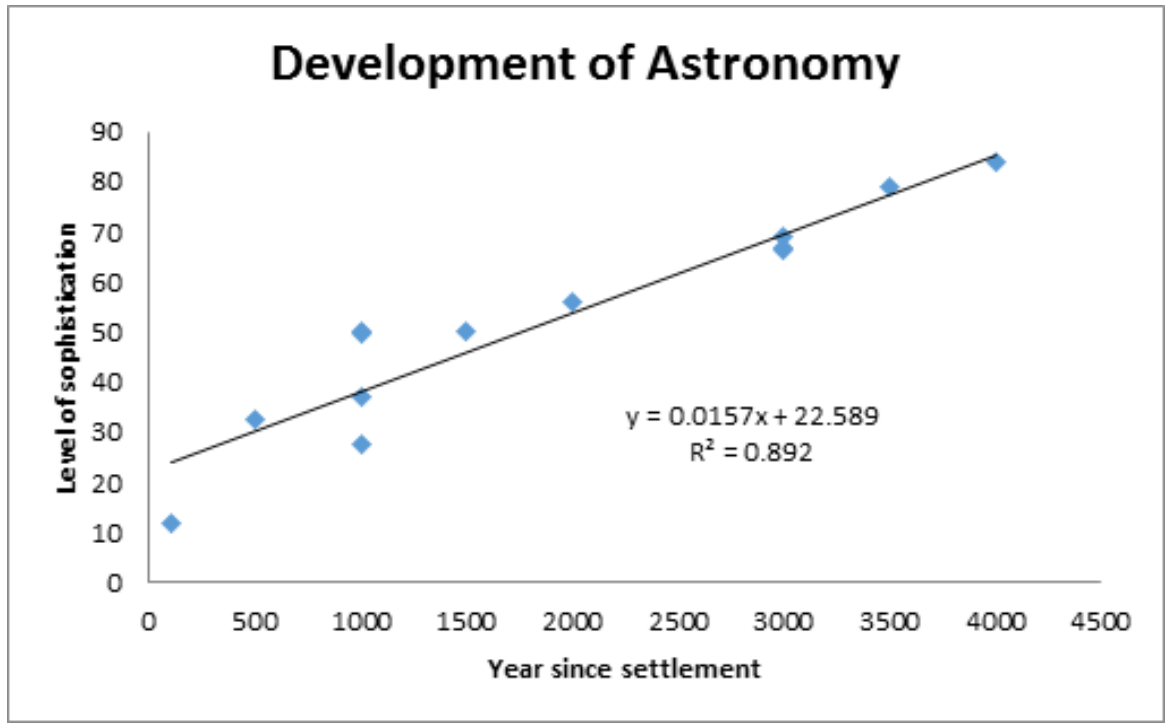

*Figure 3 Growth of complexity of civilisation with period of settlement (see table 1 for the data)*

This growth has not been without a certain cost to the culture in terms of manipulation of environment, loss of bio diversity etc. But we shall not deal with these issues here.

Figure 3 demonstrates the relation between the period of settlement and the level of complexity. As can be seen from the figure, there is a clear correlation between the two across continents and cultures.

4. Development of society

Societies grow in a complex manner typically by inventing new technologies or ideas. These technologies and ideas take time to develop and do not grow to their full potential suddenly (Vahia and Yadav, 2011 and references therein).

We present a simplified diagram of these complexities in figure 4 below. Solid line in the figure shows the potential of a scientific or technological breakthrough while the dotted line shows the level to which the potential of the technology (or idea) is explored by a society. When the society reaches the level where the maximum advantage has been derived from a given technology, it needs to invent new technologies or new ways or organisation in order to meet the increasing demands from the population for a better lifestyle.

While this stepwise increase in the capabilities with time and the gradual rise in its exploitation can be traced back to the earliest history (Vahia and Yadav, 2011), recent history itself shows some obvious examples. The early history is conventionally classified as Stone Age, Bronze Age and Iron Age.  The Iron Age itself lasted till the Second World War and it took us approximately 3500 years to fully exploit the iron technology. Beyond that it is safe to assume that we entered the silicon age with the advent of semiconductor technology. But we were far quicker in exploiting this technology and in about 60 years, we have seen its full exploitation. We are already beginning to see the saturation of this technology as we approach the end of the era where Moore's law was valid.

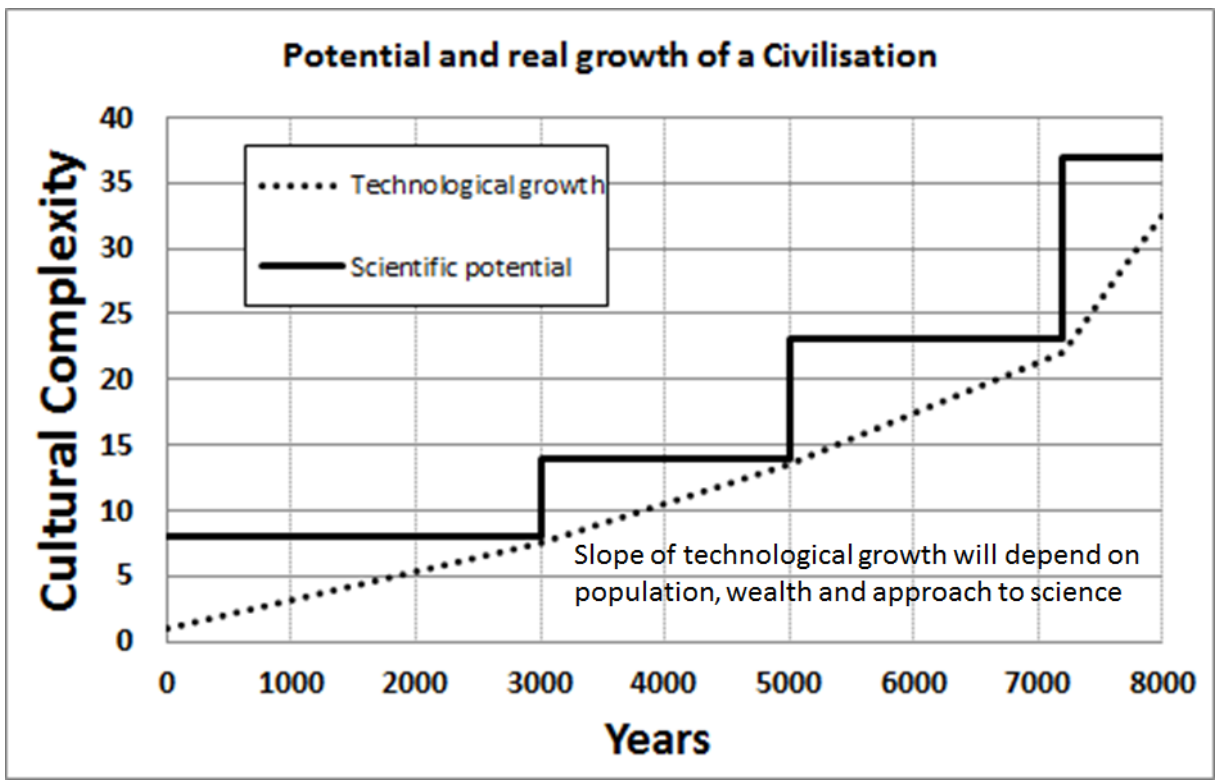

*Figure 4: Growth of technological potential with time.*

We are now probably entering the carbon age and it is impossible to predict how long it will take us to fully exploit this technology since carbon is a semiconductor and a versatile atom capable of making complex 4 atom structure and long chains that make it far more versatile than silicon whose only virtue is the control it provided on very small devices that could be manipulated by small changes in voltage. It is therefore useful to note that the age of carbon, does not imply an age of biology. While carbon is crucial to life and biology, carbon is far more versatile. With its intrinsic semiconductor property, its multiple forms from soot to diamond and its chemical versatility, it is one of the most flexible elements. Creating life giving molecules is just one of its many capabilities. The age of carbon therefore promises to be far more versatile and greater in reach than the age of silicon – which also has 4 valence bands but is far too heavy and not as versatile a receptor as carbon.

This growth of humans from being one of the animals to the most dominant has been reached through a complex interplay of streamlined brains, systematic studies, and invention of recording techniques that allow us to hold on to past experiences and build on past achievements over long periods of time.

All this results in continuing studies of nature and environment and in identifying the potentially exploitable technology for the large good of human beings. Curie and Mace (2009) have shown that the success of an ethno-linguistic group depends on its political sophistication as derived from the complexity of the political order. Gelfand et al. (2011) have studied 33 societies and show that there is a significant variation in the cultural norms that can be classified as tight (have many strong norms and a low tolerance of deviant behaviour) or loose (have weak social norms and a high tolerance of deviant behaviour) and different societies respond differently to the same stimuli based on past history.

However, even within this broad scope, each individual society will respond differently based on three major factors, namely history, geography, and experiences as well as knowledge acquired in the past. Geography in particular will decide which resources a society will value the most, history will decide on its response to any specific stimulus. Under identical stimuli and faced with identical problems, different cultures will respond differently even if two of the three parameters are similar. This is largely driven by the proximity of the people to any of the parameters and their sensitivity and recollection. History in particular is the most subjective of these parameters and the most influential. Probably the most dramatic response of the collective sense of need of a society is the response of the Soviet Union after the communist revolution, when challenged by the western neighbours with advanced science and technology forced them to adopt to science at the cost of all else. They responded by banning religion and negating history. The rise of the Soviet Union as a scientific superpower was probably one of the shortest and the fastest rise of a civilisation from one that barely tinkered with the dominant scientific cultures to a nation that could achieve all its dreams of power and dominance with its science and technology within the few decades. While, in principle, it is possible to make a similar statement about China's recent rise, there are several important differences. China has had a long tradition of science and religion has not been an important drag on its thinking. Also, unlike the Soviet Union, China has risen initially by imitation and though it is a major economic power, its capabilities in pure sciences and long lead time vision based applied sciences is also limited. Its rise is primarily attributed to technology transfer and imitation.

Conclusions

The brain has evolved from a survival aid to powerful organ in Homo Sapiens that can think and comprehend and understand the universe around it. This brain is capable of comprehending and analysing the universe. The growth has involved a complex play of experimentation and interaction as well as analysis of the environment as an independent entity that we call science. We will analyse the evolution of science in the next paper (Vahia, 2016).

References

1. Armitage S J, et al., 2011, The Southern Route "Out of Africa": Evidence for an Early Expansion of Modern Humans into Arabia, Science, 331, 453
2. Bakker R T, 1971, Dinosaur Physiology and the evolution of Mammals, Evolution, 25, 636
3. Balter, M, 2002, What made humans modern, Science, 295, 1219
4. Balter M, 2010, Report on the 11[th] conference of archaeozoology, Science, 329, 1464
5. Balter M, 2014, The killing grounds, Science, 344, 1080 - 1083
6. Batty M, 2013, The new science of cities, MIT Press Books,
7. Conard N J, 2009, A female figurine from the basal Aurignacian of Hohle Fels Cave in south Western Germany, Nature, 459, 448
8. Curie T E and Mace R, 2009, Political complexity predicts the spread of thnolinguistic groups, PNAS, 106, 7339
9. Dediu D and Ladd D R, 2007, Linguistic tone is related to the population frequency of the adaptive haplogroups of two brain size genes, ASPM and Microcephalin PNAS, 104: 10944
10. Dediu D and Levinson S C, 2013, On the antiquity of Languages: the reinterpretation of Neanderthal linguistic capacities and its consequences, Frontiers of Psychology, 4, 1
11. d'Errico F. et al., 2009, Additional evidence on the use of personal ornaments in the Middle Palaeolithic of North Africa, Publications of the National Academy of Sciences, USA, 106, 16051
12. Diamond J, 2005, Guns, germs and Steel, Vintage books,
13. Diamond Jared, 2009, Archaeology: Maya, Khmer and Inca, *Nature*, 461, 479
14. Feigenson L and Halberda J, 2008, Conceptual knowledge increases infants' memory capacity, PNAS, 105, 9926.
15. Fitch T W, 2010, The Evolution of Language, Cambridge University Press,
16. Frith Chris and Frith Uta, 2005, Current Biology, Volume 15, R644 doi:10.1016/j.cub.2005.08.041
17.
18. Gelfand M J, 2011, Differences Between Tight and Loose Cultures: A 33-Nation Study, Science, 332, 1100.
19. Guy F, et al., 2005, Morphological affinities of the Sahelanthropus tchadensis (Late iocene hominid from Chad) cranium, PNAS, 102, 18836
20. Gibbons, A, 2013, Elusive Denisovans Sighted in Oldest Human DNA, Science, 342, 1156
21. Henshilwood C A et al;2002, Emergence of Modern Human Behavior: Middle Stone Age Engravings from South Africa, Science, 295, 1278
22. Henshilwood C A et al., 2011, 100,000-Year-Old Ochre-Processing Workshop at

Blombos Cave, South Africa,  Science, 334, 219

23. Huffman Michael A., 2003, Animal self-medication and ethno-medicine: exploration and exploitation of the medicinal properties of plants. Proceedings of the Nutrition Society, 62, pp 371-381. doi:10.1079/PNS2003257.

24. Hurley M M, Dennett D C, Adams R B Jr, 2011, Inside Jokes, MIT Press, Cambridge

25. Iyengar R N, 2013, Parasaratantra: Ancient Sanskrit Texts on Astronomy and Natural Sciences, Jain University Press, 2011

26. Kielan-jaworows Zofia, 2004, Richard L Cifelli and Zhe-xi Luo, Mammals from the Age of Dinosaurs - Origins, Evolution and Structure, Columbia University Press

27. Kuzawa  Christopher W., Chugani Harry T., Grossman Lawrence I., Lipovich Leonard, Muzik Otto, Hof Patrick R., Wildman Derek E., Sherwood Chet C., Leonard  William R., and Lange Nicholas, 2014, Metabolic costs and evolutionary implications of human brain development, www.pnas.org/cgi/doi/10.1073/pnas.1323099111

28. Kuppanna Sastry T S, Vedanga Jyotisa by Lagadha (edited and translated), critically edited by K V Sarma, Indian National Science Academy, New Delhi, India, 1985.

29. Lewis-Williams D, 2002, The mind in the cave, Thames and Hudson publication,

30. Lewis-Williams D and Pearce D, 2005, Inside the Neolithic Mind, Thames and Hudson publication,

31. McBearty S, 2012, Sharpening the mind, Nature,  491, 531

32. Mithen S, The prehistory of the Mind, A Phoenix Paperback of Thames and Hudson publication, 1996.

33. Nettle D, 2007, Language and genes: A new perspective on the origins of human cultural diversity, PNAS, 2007, 104: 10755–10756

34.

35. Premack David and Woodruff Guy, 1978, Does the chimpanzee have a theory of mind?. Behavioral and Brain Sciences, 1, pp 515-526 doi:10.1017/S0140525X00076512

36. Roebroeks W and Villa P, 2012,  On the earliest evidence for habitual use of fire in Europe, Publications of the National Academy of Sciences USA, 108, 5209–5214

37. Tomasello M, Why we cooperate, 2009, A Boston Review Book

38. Vahia, M N, 2015, Current Science, , Evolution of science II: Insights into working of Nature

39.

40. Vahia M N, 2014, Physical Sciences and the Future of India, Manipal University Press.

41. Vahia M N and Yadav N, Reconstructing the history of Harappan Civilisation, Journal of Social Evolution and History, 2011, 10, 67-86. 2011

42. Vahia M N, Yadav N and Menon S, 2015, Perspectives on origin of Astronomy in India,  National Council for Science Museums,

43. Vernot B and Akey J M, Resurrecting Surviving Neandertal Lineages from Modern Human Genomes, Science, 2014, 343, 1017

44. Wadley L, et al., 2011, Middle Stone Age Bedding Construction and Settlement Patterns at Sibudu, South Africa, Science,  334, 1388